\documentclass[]{article}
\usepackage[utf8]{inputenc}
\usepackage{url}

\usepackage{graphics}
\usepackage{graphicx}

\begin{document}




\title{Beyond COVID-19 Pandemic: Topology-aware optimisation of vaccination strategy for minimising virus spreading}

\author{F. Petrizzelli, P.H. Guzzi, T. Mazza}
\date{F.P. and T.M. are within Laboratory of Bioinformatics, Fondazione IRCCS Casa Sollievo della Sofferenza; \\ P.H.G. is within Department of Surgical and Medical Sciences, University of Catanzaro, hguzzi@unicz.it}



\maketitle


\begin{abstract} 
The mitigation of an infectious disease spreading has recently gained considerable attention from the research community. It may be obtained by adopting sanitary measurements (e.g., vaccination, wearing masks), social rules (e.g., social distancing), together with an extensive vaccination campaign. Vaccination is currently the primary way for mitigating the Coronavirus Disease (COVID-19) outbreak without severe lockdown. Its effectiveness also depends on the number and timeliness of administrations and thus demands strict prioritization criteria. Almost all countries have prioritized similar classes of exposed workers: healthcare professionals and the elderly, obtaining to maximize the survival of patients and years of life saved. Nevertheless, the virus is currently spreading at high rates, and any prioritization criterion so far adopted did not account for the structural organization of the contact networks. 
We reckon that a network where nodes are people while the edges represent their social contacts may efficiently model the virus's spreading. It is known that spreading may be efficiently stopped by disconnecting the such a network, i.e., by vaccinating the most \textit{central} or relevant nodes, thereby eliminating the "bridge edges." We then introduce such a model and discuss on a topology-aware versus an age-based vaccination strategy.
\end{abstract}

\section*{Introduction}
Each successful outbreak containment strategy relies on three main pillars: detection (e.g., diagnostics tests), prevention (e.g., vaccines, containment), and cure (e.g., the existence of effective drugs). While the last measure is effective when an outbreak has happened, appropriate detection and prevention strategies that may be used at every stage of an outbreak, from the detection to the resolution, are required. In particular, the development of effective diagnostic tests is technically challenging, which are helpful to detect infected people but only limitedly applicable for spreading containment. It is thus evident that appropriate strategies for containment to ensure that future outbreaks can be more effectively contained are needed. 

Containment strategies fall into two main categories, often interlaced: (i) social and sanitary containment, which are based on the limitation of contacts among people but require many social efforts (e.g. lockdown, change of behaviour, use of protecting device such as face masks); (ii) vaccination strategies, which tackle the spreading without imposing social limitations. For instance, for the SARS-CoV-2 pandemic \cite{gostin2021coronavirus}, after initial containment measures and the trial of different therapies, many research efforts led to the development of different vaccines \cite{guo2020origin,le2020covid,kumar2021data,ortuso2021structural}. However, considering the spread of COVID-19 and the production rates, there is a need for developing ad-hoc prioritization strategies \cite{bubar2021model} that are also effective for the subsequent pandemics. Similarly, even in presence of a sufficient amount of vaccines, it is important to consider the speed of spreading that may be higher than the rate of immunisation, thus infecting the vaccination strategy. This was clear in COVID-19 third wave and this may happen even in future pandemic unfortunately. It is clear that the availability of a vaccine prioritization strategy is then a crucial challenge in fighting COVID-19 and future pandemics since there is evidence that the production rate of vaccines may remain insufficient and that viruses may have different impact and transmission rates in other social groups as demonstrated by SARS-CoV-2 related diseases \cite{khamsi2020if,galicia2020predicting,zucco2020sentiment,guzzi2020master}. 

Considering COVID-19, the evidence that the disease showed a higher fatality rate in older people \cite{cannistraci2020age,goldstein2021vaccinating}, and that healthcare workers showed a higher risk, many governments gave priority to such classes. As demonstrated by Goldstein et al. in  \cite{goldstein2021vaccinating}, this strategy enabled to save both the most lives and most years of life. Common sense suggests that a good prioritization scheme should choose the best trade-off between saving the maximum number of lives and the most future life. The mathematical model developed by the authors demonstrated that giving priority to older adults may maximize both effects. Thus, the strategy is feasible. In fact, older people's prioritization was chosen as the main criterium in many countries such as Italy and the US. These countries also gave priority to healthcare professionals, teachers, and caregivers. Despite the effectiveness of this approach, it has been clear that a vaccine allocation strategy requires the incorporation of a model of transmission and the epidemiological characteristics of the disease among social groups \cite{buckner2021dynamic,giordano2020modelling,maheshwari2020network}. For instance, Jentsch et al. \cite{jentsch2021prioritising} discussed the problem of the optimization of the strategy of vaccination. They demonstrated that a strategy based only on age is not optimal compared to a contact-based strategy.

The analysis of these approaches suggests, to the best of our knowledge, two primary considerations: (i) the optimization strategy is dependent on the desired goals, (ii) the integration of the characteristics of the modeling improves the performances, (iii) a dynamic strategy may outperform a static one. Despite this, we retain that modeling the spreading using a classical SEIR (Susceptible-Exposed-Infective-Recovered) model may not be the best choice since some parameters are considered at a global scale, while the spreading involves single contacts. In parallel, some previous works such as \cite{alguliyev2021graph,bryant2020modelling,karaivanov2020social,zaplotnik2020simulation,das2021analyzing,patil2021assessing} have demonstrated that the use of a model coming from graph theory may be helpful to describe the spreading. In this way, comprehensive graphs may be derived using nodes, i.e., people, and edges, i.e., their contacts. Hence, there is the need for the introduction of novel methods and supporting tools for the implementation of an optimal vaccination strategy. 

Consequently, we propose a mixed framework for the simulation of disease spread on networks. As a first iteration, it was designed to model the number of (i) susceptible individuals, (ii) infected individuals, and (iii) recovered individuals. This so-called SIR model was tested on networks of increasing sizes, random and natural topology configurations, thereby mimicking the social contacts at most, and were subjected to vaccination in different instants of the infection to test the efficacy and timeliness of the implemented vaccination strategy. The way vaccination was mimicked deserves, in fact, particular attention. We propose removing, i.e., to vaccine the nodes with the highest topological centrality metrics within the network while controlling the consequent epidemic spreading reduction by a linear algebra formulation. 
The application of this framework supports our initial hypothesis that any strategy of optimisation that is unaware of virus spreading topology, without severe lockdown, may fail in virus circulation mitigation. This may favourite the insurgence of novel virus variants during the vaccination that are not covered by existing vaccines.








\section*{Related Work}
\label{sec:related}
\subsection*{Epidemics Control on Network}
Frameworks based on graph theory and network science are currently largely used for modeling and studying diffusion processes in several scenarios: ideas in social networks, objects in transport networks, and virus spreading. The COVID-19 outbreak has given researchers both an unprecedented source of data and a real application scenario. In \cite{plazas2021modeling}, authors used a multiplex network and a SIR system \cite{boccaletti2014structure} to model heterogeneous contacts among humans. They considered three kinds of contacts:  a Household layer, a Work layer, and a Social layer. The SIR process mimicked the epidemic spreading. The framework's objective was to compare partition strategies that model lockdown to evaluate the control of the epidemic outbreak and minimize the economic cost associated with the partial lockdown.

Similarly to multiplex networks, temporal networks coupled to SIR model have been used to model infection spreading \cite{colizza2007reaction,zhan2020susceptible}. In \cite{humphries2021systematic}, a temporal network implemented as a multiplex network with time-varying edges was used to model an epidemic. Authors also found a condition discriminating controlled/uncontrolled epidemic on the basis of the parameters of the SIR model and on the matrix which  describes the evolution of the SIR. 


The spreading of information (or more generally the spreading itself) and best conditions for spreading and the emergence of super-spreaders has been largely investigated in network science \cite{paluch2018fast,ash2017superspreaders}. These works discussed the challenge of detecting and suppressing the spread of dangerous viruses, pathogens, and misinformation or gossip. In \cite{stegehuis2016epidemic}, authors investigated the impact of the community structure of the network on percolation, simulating the spreading of an epidemic modeled through a SIR model. Authors concluded that spreading within communities is critically related to the network denseness \cite{menniti2013,mazza2010}. At the same time, the inter-community edges are the most critical factor in spreading an epidemic, regardless of community size and shape. In particular, in \cite{wang2003epidemic} a study of epidemic spreading using the adjacency matrix of a graph has been proposed. In this work, given a contact matrix (without any constraint on the structure of the matrix) and an epidemic modeled using a SIR model, which in turn was described by two parameters $\beta$, i.e., the rate of novel infected, and $\delta$, i.e., the rate of recovered, it has been demonstrated that there is an upper bound to the epidemic given by the Eq. \ref{eq1}:
\begin{equation}\label{eq1}
 \frac{\beta}{\delta} \leq \frac{1}{\lambda_{max}}
\end{equation} where $\lambda_{max}$ is the largest eigenvalue of the adjacency matrix. Consequently, given the following property:

\begin{equation}
deg_{avg} \leq \lambda_{max} deg_{max}
\end{equation}where $deg_{avg}$ and $deg_{max}$ are the average and the maximum node degree in a network, reducing $\lambda_{max}$ would lead to a maximal network disconnection and, then, to the highest reduction of viral diffusion probability.

\section*{Results}
\subsection*{Experiment on Synthetic Networks}
\begin{figure}[ht]
    \centering
    \includegraphics[width= 4.8in]{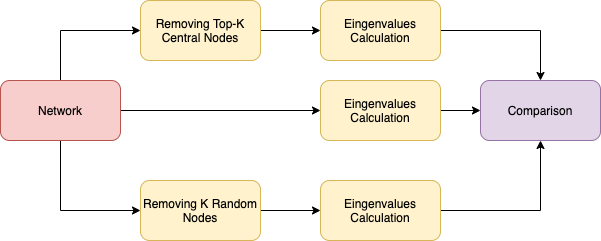}
    \caption{We initially build a test network. Then, we identified the k most central nodes. We build two networks: one in which we removed the top-k central nodes and a second one in which we removed k randomly selected node. We compared the spectra of the adjacency matrices of these two networks with respect to the spectrum of the input network. We noted that the largest eingevalues of the adjacency matrix exhibited a greater decrease when considering the removal of central nodes.}
    \label{fig:workflow}
\end{figure}

Test graphs were generated randomly according to several well-known topology characteristics: Erd\H{o}s Rényi, Random, Duplication Divergence, and Barab\'{a}si-Albert. For each corresponding adjacency matrix, we calculated the largest eigenvalue. These were calculated for the same graphs but deprived of either k random nodes or k top-scored nodes according to the following topological metrics: degree, betweenness, or eigenvalue centrality. We then reported the $\lambda_{max}$ reduction values obtained with both node elimination strategies and whether their differences were significant (Table \ref{tab:res1}). The decrease obtained when removing the topologically central nodes was higher in all cases (Figure \ref{fig:workflow}). Therefore, the impact on the ratio $1/\lambda$ was significantly higher.

\begin{table}[ht]
\caption{Largest eigenvalues (expressed as mean and standard deviation), calculated for each graph model and centrality measure (CM), of the original graph (Original), the graph randomly deprived of k nodes (Random), and the graph deprived of the k most central nodes within the graph (Top-K). DC stands for \textit{degree centrality}, BC for \textit{betweenees centrality}, EC for \textit{eigenvalue centrality} and CC for \textit{closeness centrality}. A p-value lower than 0.05 means that the decrease of the magnitudo of the largest eigenvalue is significantly higher after deleting Top-K central nodes.}
\centering
\begin{tabular}{|p{3cm}|c|c|c|c|c|} \hline 
{\bf Graph Model} & {\bf CM} &{\bf Original}  & {\bf Top-K} & {\bf Random} &{\bf p-value}   \\ \hline
Erd\H{o}s Rényi&  DC &  400 $\pm$ 0.51  & 351.2 $\pm$ 0.65  & 368 $\pm$ 0.7 & $\leq$ 0.01 \\ \hline
 Erd\H{o}s Rényi & BC &  400 $\pm$ 0.51  & 354.2 $\pm$ 0.65  & 365 $\pm$ 0.7 & $\leq$ 0.01 \\\hline Erd\H{o}s Rényi & EC &  400 $\pm$ 0.51  & 351.4 $\pm$ 0.21 & 367$\pm$ 0.92 & $\leq$ 0.01 \\  \hline \hline
 $G_{n,p}$ Random Graph & DC & 400.18 $\pm$ 0.66 & 	354.59 $\pm$ 0.61 & 	365.44
	$\pm$ 0,62 & $\leq$ 0.01 \\ \hline
	$G_{n,p}$ Random Graph & BC & 400.18 $\pm$ 0.66 & 	357.79 $\pm$ 0.61 & 	368.44 
	$\pm$ 0,62 & $\leq$ 0.01 \\ \hline 
	$G_{n,p}$ Random Graph & EC & 400.18  $\pm$ 0.66 & 	351.23 $\pm$ 0.21 & 	371.44
	$\pm$ 0.62 & $\leq$ 0.01 \\ \hline \hline 
	
	Duplication Divergence  & DC &15.2 $\pm$ 1.81 & 4.64 $\pm$ 0.52 & 15.08 $\pm$ 1.79 & $\leq$ 0.01 \\ \hline
	Duplication Divergence  & BC &15.2 $\pm$ 1.81 & 5.67 $\pm$ 0.51 & 14.88 $\pm$ 1.19 & $\leq$ 0.01 \\
 \hline
 Duplication Divergence & EC & 15.2 $\pm$ 1.81 & 8.53 $\pm$ 1.4 & 17.14 $\pm$ 1.14 & $\leq$ 0.01 \\ \hline \hline
 Barab\'{a}si-Albert  & DC & 129.38 $\pm$ 1.9 & 67.43 $\pm$ 3.45 & 115.85 $\pm$ 1.79 & $\leq$ 0.01 \\ \hline
Barab\'{a}si-Albert  & BC & 129.38 $\pm$ 1.81 & 65.79 $\pm$ 0.51 & 120.98 $\pm$ 1.19 & $\leq$ 0.01 \\
 \hline
Barab\'{a}si-Albert & EC & 129.38 $\pm$ 1.9 & 68.53 $\pm$ 1.4 & 117.14 $\pm$ 1.14 & $\leq$ 0.01 \\ \hline
\end{tabular}\label{tab:res1}
\end{table}

\subsection*{Experiments on Real Networks}
Experimenting the analytical methods presented in the previous section on the daily dynamic contact networks collected during the \textit{Infectious SocioPatterns} event that took place at the Science Gallery in Dublin, Ireland, in 2009, we confirmed that eliminating the k most central nodes caused a superior reduction of $\lambda_{max}$ than removing k random nodes. This held irrespective of the considered centrality metrics (Table \ref{tab:resreal}).
\begin{table}[ht]
\caption{Largest eigenvalues (expressed as mean and standard deviation), calculated on the for each graph model and centrality measure (CM), of the original graph (Original), the graph randomly deprived of k nodes (Random), and the graph deprived of the k most central nodes within the graph (Top-K).}
\centering
\begin{tabular}{|p{3.3cm}|c|c|c|c|c|} \hline 
{\bf Graph} & {\bf CM} &{\bf Original}  & {\bf Top-K} & {\bf Random} &{\bf  p-value}   \\ \hline
Infectious Graph Dublin &  DC &  23.0 $\pm$ 0.21 & 14.35 $\pm$ 0.0 & 23.36 $\pm$ 0.7 & $\leq$ 0.01 \\ \hline
 Infectious Graph Dublin& BC &   23.0 $\pm$ 0.21 & 16.76 $\pm$ 0.0 & 23.36 $\pm$ 0.7 & $\leq$ 0.01 \\\hline Infectious Graph Dublin & EC &  23.0 $\pm$ 0.21 & 16.88 $\pm$ 0.0 & 23.36 $\pm$ 0.7 & $\leq$ 0.01  \\  \hline \hline

\end{tabular}\label{tab:resreal}
\end{table}

\subsection*{Speeding-up Herd Immunity}
Herd immunity is the {\it Holy Grail} of all the people that are fighting against COVID-19. Herd immunity is a kind of indirect protection from an infectious disease that may happen when a significant fraction of the population has become immune to infection. Individuals become immune by recovering from an earlier infection or through vaccination. 

For COVID-19, the herd immunity was deeply investigated \cite{aschwanden2021five,fontanet2020covid,randolph2020herd}, focusing on possible effects that may hamper herd immunity, e.g., number of infected people, time, the insurgence of variants that may disrupt the herd immunity equilibrium. Generally, all agree that herd immunity can be reached when 60\%-70\% of people become immune. Thus, supposing that people become immune only for the effect of vaccination, we eliminated $n_{h}=$70\% of nodes from our networks and calculated the largest eigenvalues, $\lambda_{H}$. We also calculated the number of central nodes $n_{hs}$ that should be removed to obtain the same largest eigenvalues. We retain that the difference $n_{hs} \leq n_{h}$ result in a lower time for reaching herd immunity confirming the strength of the approach. Table \ref{tab:herdimmunity} summarizes these results.

\begin{table}[ht]
\caption{The number of people, $n_{hs}$, which should be vaccinated following the optimized strategy for reaching herd immunity}
\label{tab:herdimmunity}
\centering
\begin{tabular}{|p{2.5cm}|c|c|c|} \hline 
{\bf Graph Model} & {\bf CM} & {\bf $n_{hs}$ } & {\bf $n_{h}$ }   \\ \hline
Erd\H{o}s Rényi&  DC & 41.1\%& 70\% \\ \hline
Erd\H{o}s Rényi&  BC & 42.5\%& 70\% \\ \hline
Erd\H{o}s Rényi&  EC & 43.1\%& 70\% 
\\ \hline \hline
\end{tabular}
\end{table}

\section*{Discussion}
The theoretical model we developed suggests that at least one rationale exists that can allow us to optimize the vaccination strategy against SARS-CoV-2. Before this study, several other independent studies have used mathematical modeling to explore new prioritization strategies. These models, as introduced before, vary widely in terms of considered populations, the model used, interventions, and assumptions \cite{medo2020contact,pare2020modeling}. However, they all agree with us that optimising the vaccination strategy may positively impact the outcomes. We proposed designing an optimization strategy based on a topology model to accomplish this aim. It was based on the assumption that the contact network was trackable in time among people and then that the most "social" people were the first targets of a national vaccination campaign. 

Our analysis was based on a simple graph-based model (SIR) built on the assumption that the perfect mixing strategy is too large since contacts among people generate a social network representable through models different from the perfect mixing. Contacts do not have geographical constraints and neglect any social aspect that may affect the network topology, in fact, locally. Moreover, we also assumed that vaccination could block both disease and transmission, even though some evidence seems to indicate that vaccines may have different efficacy in blocking transmission \cite{mallapaty2021can}.

We are also aware that our model would benefit from the availability of contact tracing data. Unfortunately, high-resolution mobility data are not currently available to the scientific community, despite the evidence of the introduction of technologies able to infer information about contacts that combine data gathered from phones in a privacy preserving way. If this was the case, it should be noted that our model would incredibly improve its performance by just enriching the edges of weights.

\begin{figure}[ht]
    \centering
    \includegraphics[width= 4.8in]{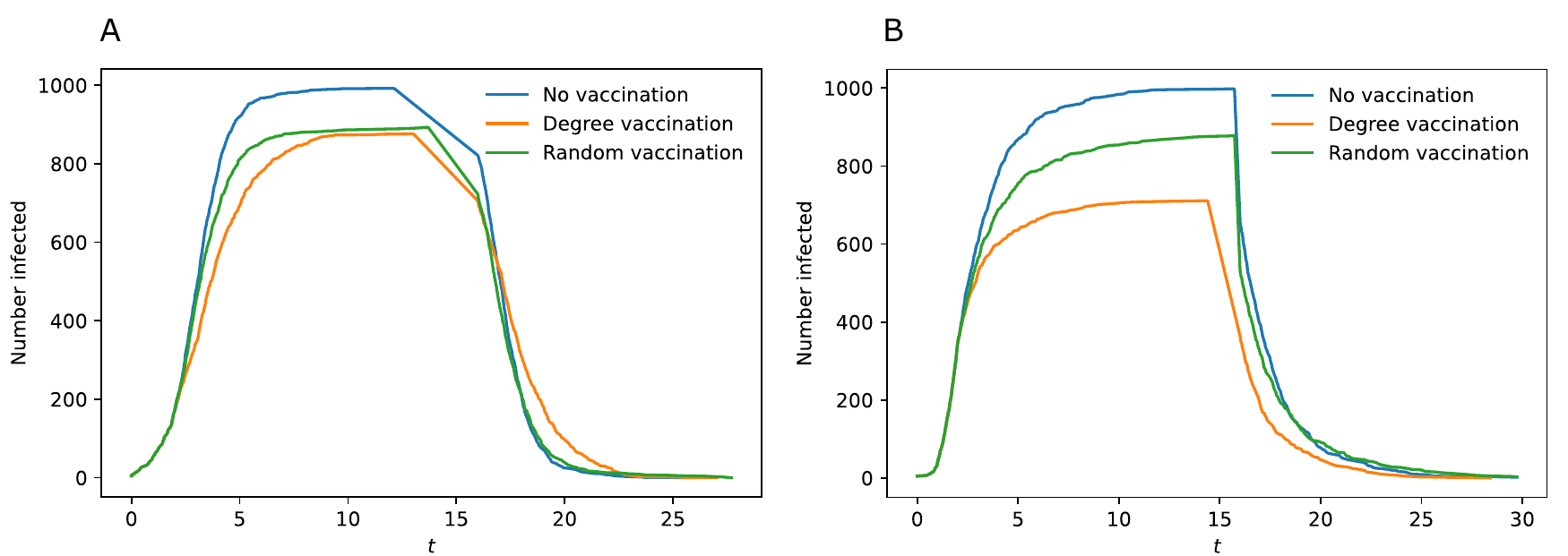}
    \caption{Impact of different vaccination strategies on the spreading of the infection using two different network models, the Erd\H{o}s-Rényi graph (A) and the Duplication Divergence generative model (B). At simulated time $t=2$, both a random vaccination (green) and a degree-based vaccination (orange) were applied, removing 100 random nodes and the $100$ top-degree nodes, respectively. As a control, the spreading of the infection without any vaccination was simulated (in blue). We observed how removing the top-degree nodes from the graph results in a rapid decrease in the number of infections compared to both control and random vaccination, especially in the Duplication Divergence Model network. At $t=14$, we can observe the rapid decrease of the three lines, with a large part of the infected population starting to recover.}
    \label{fig:sim}
\end{figure}

Our work showed how the implemented optimization strategy could provide good results for all the centrality metrics and our studied network structures. These results were significantly better than those obtained with the same methods and models but applying a random vaccination plan (an example in Figure \ref{fig:sim}). Moreover, we showed how the largest eigenvalue could be effectively associated with topological metrics other than the node degree centrality. This leaves room for the possibility to improve vaccination performance further using slightly more complex, still intuitive, topological metrics, like the Borgatti's \textit{group centrality} or the \textit{keyplayer} matrics \cite{borgatti1999, borgatti2006}, as implemented in Pyntacle \cite{mazza2020}.

Herd immunity was considered achievable with at least 70\% of vaccinated people. In this simplified setting, we further showed how a vaccination strategy driven by a topological screening might help reach it in a shorter time. 

\section*{Methods}
\label{sec:methods}
\begin{figure}[ht]
    \includegraphics[width= 4.8in]{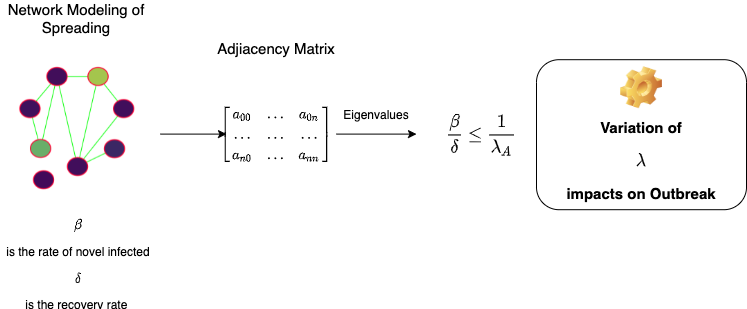}
    \caption{The rationale of the paper. The spreading of the SARS-CoV-2 may be modeled using a network whose nodes are people while edges contacts among them. The spreading may be summarised in a SIR model using two main parameters $\beta$, and $\delta$, representing the rate of novel infected people and the recovery rate, respectively. The network may be represented using the adjacency matrix of the resulting graph. It is known that spreading may be contained when the following condition holds: $\frac{\beta}{\delta}$ $\leq$ $\frac{1}{\lambda}$, where $\lambda$ is the largest eigenvalue of the adjacency matrix. We hypothesize that an individual's vaccination is equivalent to the deleting of a node (or equivalently to the deleting of the edge of the contacts).}
    \label{fig:example}
\end{figure}

We considered four different models of random networks: \begin{itemize}
    \item Erd\H{o}s-Rényi
    \item Duplication Divergence
    \item Geometric-Random
    \item Barab\'{a}si-Albert
\end{itemize} 
We generated 100 randomized networks with the same degree distribution and shuffled edges for each network model. For each node of the network, we calculated the degree, betweenness, closeness, eigenvector centrality, and, finally, the eigenvalues of the adjacency matrix. We removed the top 100 nodes having the highest value for each of these metrics and repeated the eigenvalues' calculation. We also calculated the eigenvalues of the graph's adjacency matrix obtained by removing 100 nodes selected at random to build a null model (Figure \ref{fig:example}).

\subsection*{Data}
\paragraph{$G_{n,p}$ model}
We generated a stochastic random graph, also known as \textit{binomial graph}, using the fast generator provided by the Python NetworkX library \cite{kumar2000stochastic}. We set the number of nodes equal to 1000 and a probability $p=0.4$ of having an edge connecting two nodes.

\paragraph{Erd\H{o}s-Rényi model}
We generated an Erd\H{o}s\-Rényi graph having 1000 nodes and a probability $p=0.4$ of establishing an edge between two nodes \cite{batagelj2005efficient}.

\paragraph{Duplication Divergence model}
We generated a random graph having the duplication divergence structure described by Ispolatov et al. in  \cite{ispolatov2005duplication}. The library created a graph of 1000 nodes  by duplicating the initial nodes and retaining edges incident to the original nodes with a retention probability of $p=0.4$.

\paragraph{Barab\'{a}si-Albert model}
We generated a Barab\'{a}si-Albert graph, a random graph built according to the Barab\'{a}si–Albert preferential attachment model \cite{barabasi1999emergence}. A graph was grown by attaching new nodes, each with 50 edges, which were preferentially attached to existing high degree nodes.

\paragraph{Random Geometric model}
We generated a random geometric graph, where 1000 nodes were placed uniformly at random in the unit cube. Two nodes were joined by an edge if their distance was at most radius \cite{penrose2003random}.

\subsection*{Real network}
The dataset of real networks considered in this study was downloaded from the Network Repository \cite{nr}. It contained the daily dynamic contact networks collected during the \textit{Infectious SocioPatterns} event that took place at the Science Gallery in Dublin, Ireland, in 2009 during the artscience exhibition INFECTIOUS: STAY AWAY. Each file in the downloadable package contains a tab-separated list representing the active contacts during 20-second intervals of one day of data collection.

\subsection*{Network Centrality Measures}
\label{sec:centralitymeasures}
As discussed before, a graph or network of $n$ nodes can be represented as an adjacency matrix ($\mathcal{A} \in \Re^{n\times n}$), where each entry in the matrix $\mathcal{A}_{ij}\neq 0$ indicates the existence of an edge between nodes $i$ and $j$, while $\mathcal{A}_{ij} = 0$ indicates the absence of an edge between the two
nodes. A special case of graphs, called {\bf edge-weighted graphs}, is characterized by an adjacency matrix whose values are real-valued values. The following discussion is focused on undirected, non-weighted graphs and may be easily extended to both ordered and edge-weighted graphs.


In the case of an unweighted network, a \emph{geodesic} (or shortest path) from node $w_i$ to node $w_j$ is the path the involves the minimum number of edges. Consequently, we may define the \emph{distance} between nodes $w_i$ and $w_j$, where $\rho_\mathbf{g}(w_i,w_j)$ is the number of edges involved in a geodesic  between $w_i$ and $w_j$. Starting from the computation of distance, a set of \emph{centrality} measures have been introduced. The aim of such measures is to evidence the relevance, or importance, of a node in a network by analyzing its local topology properties.

\subsubsection*{Degree centrality}
The \textbf{degree centrality} of a node $v_i$ is the number of its adjacent nodes.
\[
C_{deg}(v_i)=deg(v_i).
\]
Sometimes, the degree centrality is normalized by the maximum degree possible of a node:
\[
C_{degnorm}(w_i) =deg(v_i) \frac{d_i(\mathbf{g})}{n-1}.
\]
The degree centrality gives some information related to the immediate relevance of a node $v_i$, but it misses some aspects of the whole structure of the network as well as the node's position. 

\subsubsection*{Closeness Centralities}
The \textbf{closeness centrality} considers the distance among nodes. Formally, the closeness centrality of a node $v_i$ is the reciprocal of the average shortest path distance to $v_i$ over all $n-1$ reachable nodes, i.e.
\[
C_{closeness}(v_i) = \frac{n-1}{\sum_{j=1}^{j=n-1,j \neq i}d(v_i,v_j)}.
\]
where $d(v_i,v_j)$ is the shortest distance among the considered pair. 

\subsubsection*{Betweenness Centrality}
While the closeness centrality indicates how close a node is to the others, \textbf{Betweenness centrality} \cite{brandes2007centrality} evaluates how much a node stands between each other. For each node pairs, $v_i$ and $v_j$, in a network, it scores a node based on the number of shortest paths passing through it and all other geodesics connecting $v_i$ and $v_j$ not passing through it. Formally, the betweenness centrality of a node $(v_i)$ is calculated as:
\[
C_{betweennes}(v_i) = \sum_{i \neq j \neq k} \frac{ \sigma_{j,k}(i)}{\sigma_{j,k}}.
\]
where, $\sigma_{j,k}$ is the total number of shortest paths from node $v_j$ to node $v_k$ and $\sigma_{j,k}(i)$ is the number of those paths that pass through $i$.

\subsubsection*{Eigenvector centrality}
Eigenvector centrality \cite{bonacich1972} scores all nodes of a network on the assumption that connections to high-scoring nodes contribute more to the score of the node rather than connections to low-scoring nodes. 
Given the adjacency matrix $\mathcal{A}$ of a graph, the eigenvector centrality of a node $v$ can be defined as:
$$ x_{v} = \frac{1}{\lambda}\sum_{w \in Neigh(v)} x_w =  \frac{1}{\lambda}\sum_{w \in \mathcal{G}} \mathcal{A}_{v,w} x_w $$
Let $x_i$ be the eigenvector centrality of $i^{th}$ node $v_i$, then $X = (x_1, x_2, \cdots x_n)^T$ the solution of equation $\mathcal{A}X = \lambda X$, where $\lambda$ is the greatest eigenvalue of $\mathcal{A}$ to ensure that all values $x_i$ are positive~\cite{okamoto2008ranking} by the Perron-Frobenius theorem. The $v^{th}$ component of the related eigenvector will give the relative centrality score of the vertex $v$ in the network.

\subsection*{Statistics}
In order to assess the significance of the differences obtained with the two implemented vaccination policies, i.e., random and based on the k top-central nodes, we used a Student's t-test for paired values. A p-value was considered significant if lower than 0.05.

\subsection*{Simulation framework implementation}
As anticipated, in this work we relied on the classical SIR model for spread of disease. It is one the simplest models for the epidemic spreading simulation, with each individual that can be in a susceptible (S), infectious (I), or recovered (R) state. Once a susceptible individual comes into contact with an infected one, it gains a probability of becoming infectious. Each infectious person can infect a susceptible neighbor and recover after a variable or fixed time span. Furthermore, recovered do not play any further role in the simulation and include both immune or deceased individuals.

To simulate the spreading of the SARS-CoV-2 virus, we implemented this model using the Python library \textit{Epidemics on Networks} (EoN) \cite{miller2019}. We simulated each of the above-described networks ten times and then average results. At the beginning of these runs, the entire population was in the \textit{S} state, with an infected group composed of $5$ individuals. Then, we set a constant transmission rate ($\tau = 0.4$) with a fixed recovery time of $14$ days to simulate the spreading. 

To test our different vaccination approaches, we stopped the simulations at different simulated times and thus removed a constant fraction of $S$ nodes based on topological considerations. As control, we ran simulations with a random vaccination strategy, which removed $100$ nodes selected randomly at specific time \textit{t}, and others without any vaccination strategy implemented. Finally, we restarted the simulations and evaluated the spreading of the infection by measuring the number of infected individuals over time. 

\section*{Conclusion}

Before COVID-19 pandemics were only an argument of study and pianification, under the hypotheis that they the probability of occurring was very very low. Unfortunately, COVID-19 has shown that some worst scenarios may happen. Therefore the scientific community has to develop novel tools to support decision maker to control and stop the spreading. Current tools include sanitary measurements (e.g., vaccination, wearing masks), social rules (e.g., social distancing), and extensive vaccination campaign.
The effectiveness of vaccination depends on the number and timeliness of administrations and thus demands strict prioritization criteria. Prioritization need appropriate models of simulation that are based on the joint use of network theory and mathematical modelling of spreading. We have shown  that a network where nodes are people while the edges represent their social contacts may efficiently model the virus's spreading. We presented some experimental evidences and a supporting tool that evidence that spreading may be efficiently stopped by disconnecting the such a network, i.e., by vaccinating the most \textit{central} or relevant nodes, thereby eliminating the "bridge edges."

\newpage 
\section*{Figures}

\newpage

\section*{Competing interests}
The authors declare that they have no competing interests.

\section*{Author's contributions}
PHG. conceived the experiments and drafted the manuscript; FP. performed the simulations; TM analysed the results and contributed drafting the manuscript. All authors wrote ,  reviewed, and approved the manuscript. 

\section*{Acknowledgements}
Italian Ministry of Health (Ricerca Corrente 2022–2025); ‘5 × 1000’ voluntary contribution.
\vfill

\bibliographystyle{bmc-mathphys}
\bibliography{bmc_article}

\end{document}